\documentclass{aa}  

\usepackage{color}
\usepackage{graphicx}
\usepackage{lscape}
\usepackage{natbib}
\usepackage{natbib,twoopt}
\usepackage[varg]{txfonts}
\usepackage{url}
\usepackage{xspace}
\usepackage{amssymb}

\usepackage[colorlinks = true,
linkcolor= blue,
citecolor = blue,
filecolor = blue,
urlcolor = blue,
]{hyperref}

\bibpunct{(}{)}{;}{a}{}{,}

\usepackage{graphicx}
\usepackage{txfonts}
\usepackage{color}
\def\Teff{$T\rm_{eff}$}
\def\Tcond{$T\rm_{cond}$}
\def\kms{$\mathrm{km\, s^{-1}}$}

\def\Vt{$V_{\rm t}$}
\def\logg{$\log\,g$}
\def\vsini{{\sl v~sin~i~}}

\def\Mv{$M_{V}$}
\def\loggf{$\log\,gf$}
\def\Rg{$R_{\rm G}$}

\def\kms{km s$^{-1}$}

\def\ElH{$\mathrm{[El/H]}$}
\def\II{\,{\sc ii}}
\def\I{\,{\sc i}}

\newcommand{\Lsol}{$L_\odot$}
\newcommand{\Msol}{$M_\odot$}

\begin{document}

\title{Lithium Cepheid V708 Car with an unusual chemical composition}

   \author 
          {
         V. V. Kovtyukh\inst{1,2}
          \and
           S. M. Andrievsky\inst{1,3}
          \and
          K. Werner\inst{2}
          \and
         S. A. Korotin\inst{4}
          \and
           A. Y. Kniazev \inst{5,6}}
           
  \institute{Astronomical Observatory, Odessa National University, 
             Shevchenko Park, 65014, Odessa, Ukraine \\
               \email{vkovtyukh@ukr.net}
         \and
 Institut f\"{u}r Astronomie und Astrophysik, Kepler Center for 
 Astro and Particle Physics, Universit\"{a}t T\"{u}bingen, 
 Sand 1, 72076 T\"{u}bingen, Germany 
         \and
 GEPI, Observatoire de Paris, Universit\'e PSL, CNRS, 
  5 Place Jules Janssen, F-92190 Meudon, France 
         \and
 Physics of stars department, Crimean Astrophysical Observatory, 
 Nauchny 298409, Republic of Crimea 
         \and
 South African Astronomical Observatory, PO Box 9, 7935 Observatory, Cape  Town, South Africa 
         \and 
 Southern African Large Telescope Foundation, PO Box 9, 7935 Observatory, Cape Town, South Africa 
  }

\date{Received date; accepted: date}

\abstract
{}
{The purpose of this work is to spectroscopically analyse the classical Cepheid V708~Car.
 A preliminary check of the spectrum of V708 Car showed that this is a lithium-rich supergiant. We also found that V708~Car has an unusual chemical composition in that the abundances of various elements correlate with their condensation
temperatures. We tried to find an explanation of this feature, which is unusual for classical Cepheids.}
{For the spectroscopic analysis, we used methods based on the assumption of local  and non-local thermodynamic equilibrium.}
{We determined the fundamental parameters of our program star V708~Car. 
This long-period Cepheid has a mass of about 12 M$_{\odot}$.
We derived the abundances of 27 chemical elements in this star. They are clearly correlated with their condensation
temperature: the higher the condensation temperature, the lower
the abundance (there are exceptions for sodium and barium, however).
We explain this peculiar chemical composition of the V708~Car
atmosphere by the gas--dust separation in the envelope of this star.
A similar mechanism leads to the observed peculiarities of the
chemical composition of $\lambda$~Boo, W~Vir, and asymptotic
giant branch stars.}
{}


\authorrunning{Kovtyukh et al.}
\titlerunning{Lithium Cepheid V708 Car}

\maketitle

\section{Introduction}

Classical Cepheids are pulsating supergiant stars with spectral classes from F to K.
They pulsate in radial mode(s). The pulsational instability of a Cepheid 
occurs when the star enters the so-called instability strip (IS) in the Hertzsprung--Russell (HR)
diagram before and after a previous red giant stage. The progenitors of Cepheids 
in the main-sequence (MS) stage are O-B type stars, whose masses range from approximately 
3 to 20 M$_{\odot}$ or more.

V708~Car was classified as a classical Cepheid with a pulsation period $P=51.4$~d
(ASAS project; \citealt{Pojmanski2002}). Later, \cite{Berdnikov2010}
confirmed this classification. V708~Car is a poorly investigated Cepheid. 
The SIMBAD database lists only eight publications related to the study of this star. These papers are mainly 
devoted to photometric studies. \cite{Berdnikov2010} also performed a
search for evolutionary changes in V708~Car using the Harvard photographic plate collection 
and CCD observations. He found that the pulsational period increases by 
about 52~s/yr.

V708~Car is a rather massive Cepheid. From its position in the HR diagram (see discussion below), 
its mass should be about 12\,\Msol, while its bolometric absolute 
magnitude is $-6.03$~mag. Based on the period--luminosity relation by \cite{Breuval2022}, the luminosity 
of the star is about 18\,400 \Lsol\ (using the correction for [Fe/H]= --0.40).

\section{Observations and data reduction}


   Spectral observations of V708\,Car were made on 2016 June 8 at a pulsation phase of 0.216
  with the  High Resolution Fibre \'echelle  Spectrograph (HRS)
  \citep[HRS;][]{2008SPIE.7014E..0KB,2010SPIE.7735E..4FB,2012SPIE.8446E..0AB,2014SPIE.9147E..6TC}
  at the Southern African Large Telescope \citep[SALT;][]{2006SPIE.6267E..0ZB,2006MNRAS.372..151O}.
  The HRS is a thermostabilised double-beam \'echelle spectrograph whose entire optical part
  is housed in vacuum to reduce temperature and mechanical influences.
  The blue arm of the spectrograph covers the spectral range of 3735--5580~\AA,
  and the red arm covers the spectral range of 5415--8870~\AA.
  The spectrograph can be used in low-resolution (LR, with a resolving power $R\approx14\,000-15\,000$),
  medium-resolution (MR, $R\approx36\,500-39\,000$), and high-resolution (HR, $R\approx67\,000-74\,000$) modes
  and is equipped with two fibers (object and sky fibers) for each mode.
  During the observations of V708\,Car, the MR mode was used
  with fibers of 2\farcs23 in diameter. The CCD detectors for the blue and red arms
  were used with a binning of 1$\times$1.
  
  The primary HRS data reduction was performed automatically using the
  SALT standard pipeline \citep{2010SPIE.7737E..25C} 
 , and the subsequent \'echelle data reduction was made using the HRS pipeline, as 
  described in detail in \citet{2019AstBu..74..208K}.

\section{Atmosphere parameters of V708 Car and its chemical composition}

\begin{figure}
 \resizebox{\hsize}{!}{\includegraphics{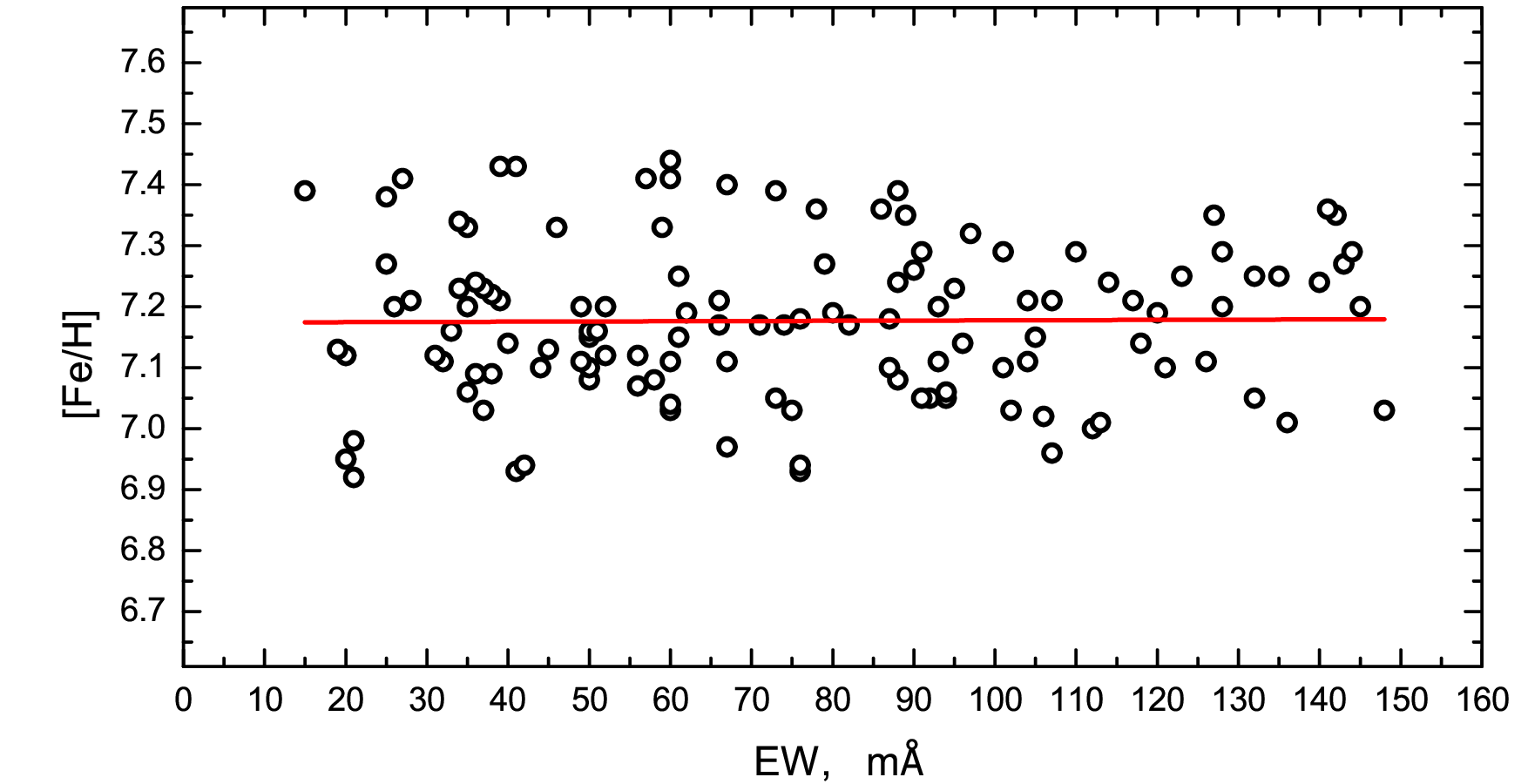}}
 \caption{ \Vt\ determination (absence of a dependence between the iron abundance 
 of individual lines and their equivalent widths). }
 \label{Vt}
 \end{figure}

The effective temperature \Teff\ of V708~Car was derived from the line-depth ratios \citep{Kovtyukh2007}. 
This technique is commonly used in studies of Cepheid variables (e.g. \citealt{Andrievskyetal2016},
\citealt{Luck2018}, \citealt{Lemasleetal2018}, \citealt{daSilvaetal2022}, \citealt{Kovtyukhetal2022}). 
When \Teff\ was determined, the surface gravity (\logg) was found by imposing the iron ionisation 
balance (the same iron abundance as derived from the neutral and ionised lines). The microturbulent 
velocity \Vt\ was derived assuming that there is no dependence between the iron abundance obtained from the 
Fe\I\ lines and their equivalent widths (EWs; Fig.\ref{Vt}). The adopted abundance value [Fe/H] was 
derived from the Fe\I\ lines because we assumed ionisation balance and because they outnumber the 
Fe\II\ lines. The atmospheric parameters \Teff, \logg, and \Vt\ are listed in Table~\ref{parameters},
together with some other parameters of the program star.

\subsection{Local thermodynamic equilibrium results}
 
 The abundances of different elements were derived in the local thermodynamic equilibrium (LTE) approximation using the atmosphere model  computed with {\sc ATLAS9} cîde  by \cite{CastelliKurucz2004} for the atmosphere parameters. 
The oscillator strengths, \loggf, were adopted from the Vienna Atomic Line Database 
({\sc VALD}, \citealt{Ryabchikovaetal2015}, version 2023). The reference solar abundances were taken from \cite{Asplundetal2009}. 

 \subsection{Non-local thermodynamic equilibrium results}

For eight chemical elements, we applied the non-LTE (NLTE) approximation to derive their abundances.
The atomic models that were used are described in detail in several papers,  
for example, carbon (\citealt{Andrievskyetal2001}, \citealt{Lyubimkovetal2015}), sodium
(\citealt{KorotinMishenina1999}, \citealt{Dobrovolskasetal2014}), magnesium
(\citealt{Misheninaetal2004}, \citealt{{Cerniauskasetal2017}}), aluminum
(\citealt{Andrievskyetal2008}, \citealt{Caffauetal2019}), sulfur (\citealt{Korotin2009}, and barium (\citealt{Andrievskyetal2009}). 

The general method of the NLTE calculations is the following. In order to find atomic level populations 
for the ions of interest, we employed the code {\sc MULTI} (\citealt{Carlsson1986}).
This program was modified by \cite{Korotin1999}. 
MULTI allows us to calculate a single NLTE line profile. When the line 
of interest was blended, we performed the following procedure. With the help
of~{\sc MULTI}, we first calculated the departure coefficients for the atomic levels 
that cause the formation of the considered line. After this,  we included 
these coefficients in the LTE synthetic spectrum code {\sc SynthV} (\citealt{Tsymbaletal2019}). 
This allowed us to calculate the source function and opacity for each studied line. 
Simultaneously, the blending lines were calculated in LTE with the help of the line list
and corresponding atomic data from the {\sc VALD} database (\citealt{Ryabchikovaetal2015}) 
in the wavelength range of the line under study. For all our computations, we 
used 1D LTE atmosphere models computed with the {\sc ATLAS9} code by \cite{CastelliKurucz2004}.
In Tables \ref{abundances} and \ref{NLTE}, we list the resulting LTE and NLTE elemental 
abundances in V708~Car.

\begin{table*}
\caption{Moment in time of the spectroscopic observation and model parameters of V708 Car.}
\label{parameters}
\centering
\begin{tabular}{ccccccccccccc}
\hline\hline   
   JD         &   P, days & phase & \Teff, K  & \logg  & \Vt, km~$s^{-1}$ &[Fe/H] \\
\hline
  2457547.738 & 51.403084 & 0.216 &  5206     & 1.30   &    4.30          & $-$0.40 \\ 
\hline
\end{tabular}
\end{table*}

\begin{table}
\begin{center}
\caption{LTE elemental abundances in V708 Car.}
\label{abundances}
\begin{tabular}{lllrrr}
\hline\hline               
  Ion  &  [El/H] & $\sigma$ & NL& (El/H)& \Tcond, K   \\
\hline
  Li\I &    +0.79&  0.20 &    1  &  1.95 &  1142    \\      
   C\I &  $-$0.32&  0.11 &    1  &  8.36 &    40    \\  
   O\I &    +0.08&  0.13 &    2  &  9.06 &   180    \\
  Na\I &    +0.21&  0.20 &    1  &  6.53 &   958    \\
  Mg\I &  $-$0.28&  0.20 &    1  &  7.40 &  1354    \\
  Al\I &  $-$0.29&  0.31 &    3  &  6.01 &  1653    \\
  Si\I &  $-$0.31&  0.08 &   14  &  7.29 &  1354    \\
   S\I &    +0.10&  0.17 &    2  &  7.31 &   664    \\
  Ca\I &  $-$0.60&  0.06 &    6  &  5.78 &  1517    \\
 Sc\II &  $-$0.87&  0.11 &    5  &  2.38 &  1659    \\
  Ti\I &  $-$0.66&  0.26 &   12  &  4.37 &  1582    \\
  Ti\II&  $-$0.93&  0.13 &    2  &  4.11 &  1582    \\
   V\I &  $-$0.68&  0.10 &    3  &  3.43 &  1429    \\
  V\II &  $-$0.55&  0.02 &    2  &  3.57 &  1429    \\
  Cr\I &  $-$0.54&  0.07 &   11  &  5.22 &  1296    \\
  Cr\II&  $-$0.25&  0.08 &    4  &  5.51 &  1296    \\
  Mn\I &  $-$0.29&  0.21 &    6  &  5.25 &  1158    \\
  Fe\I &  $-$0.40&  0.12 &  141  &  7.18 &  1334    \\
 Fe\II &  $-$0.39&  0.10 &   16  &  7.19 &  1334    \\
  Co\I &  $-$0.59&  0.16 &    4  &  4.40 &  1352    \\
  Ni\I &  $-$0.45&  0.17 &   30  &  5.85 &  1353    \\
  Cu\I &  $-$0.33&  0.16 &    2  &  3.96 &  1037    \\
  Zn\I &    +0.34&  0.21 &    1  &  4.79 &   726    \\
  Y\II &  $-$0.72&  0.02 &    4  &  1.47 &  1659    \\
 Zr\II &  $-$1.05&  0.22 &    1  &  1.83 &  1741    \\
 La\II &  $-$0.47&  0.13 &    6  &  0.86 &  1578    \\
 Ce\II &  $-$0.56&  0.22 &    6  &  1.15 &  1478    \\
 Pr\II &  $-$0.81&  0.14 &    3  &  0.00 &  1582    \\
 Nd\II &  $-$0.57&  0.20 &    7  &  0.95 &  1602    \\
 Sm\II &  $-$0.42&  0.11 &    3  &  0.68 &  1590   \\
 Eu\II &  $-$0.18&  0.22 &    1  &  0.78 &  1356    \\
\hline
\end{tabular}
\end{center}
\begin{itemize}
\item[] {\it Remarks:}

NL is the number of lines used,\\ 
 \ElH\ is the abundance relative to the solar value,\\
(El/H) is the absolute abundance value on a scale where the hydrogen abundance is 12.00,\\ 
\Tcond\ is the condensation temperature.
\end{itemize}
\end{table}

\begin{table}
\begin{center}
\caption{NLTE elemental abundances in V708 Car.}
\label{NLTE}
\begin{tabular}{llccrrr}
\hline\hline  
   Ion& (El/H)&   $\sigma$& NL & [El/H] &  Sun & \Tcond, K   \\  
\hline                       
  C\I  & 8.10  &    0.12  &   5&$-$0.33 &   8.43 &   40  \\
  O\I  & 8.88  &    0.10  &   4&  +0.16 &   8.71 &  180  \\
 Na\I  & 6.29  &    0.05  &   6&  +0.04 &   6.25 &  958  \\
 Mg\I  & 7.03  &    0.15  &   5&$-$0.51 &   7.54 & 1354  \\
 Al\I  & 5.90  &    0.20  &   3&$-$0.53 &   6.43 & 1653  \\
  S\I  & 7.10  &    0.12  &   4&$-$0.06 &   7.16 &  664  \\
 Cu\I  & 4.05  &    0.10  &   4&$-$0.20 &   4.25 & 1037  \\
 Ba\II & 2.34  &    0.10  &   4&$-$0.17 &   2.17 & 1455  \\
\hline
\end{tabular}
\end{center}
\begin{itemize}
\item[] {\it Remarks:}

For the Sun, our NLTE abundaces are listed.
\end{itemize}
\end{table}

The star is not so far away from the Sun. Its heliocentric distance is only about 4.4\,kpc \citep[\emph{Gaia} EDR3, ][]{2021A&A...649A...1G}.  
The same distance results from the period--luminosity relation  (\citealt{Wangetal2018}, see \citealt{Skowronetal2019} for details). 
The galactocentric distance is 8\,kpc. Therefore, the decreased iron abundance in this star could mean that it is a thick-disc Cepheid, 
but its distance below the Galactic plane is only $-$185\,pc, which means that it could also be a member of the thin-disc population.
 
\subsection{Lithium in V708 Car}

\begin{figure}
\resizebox{\hsize}{!}{\includegraphics{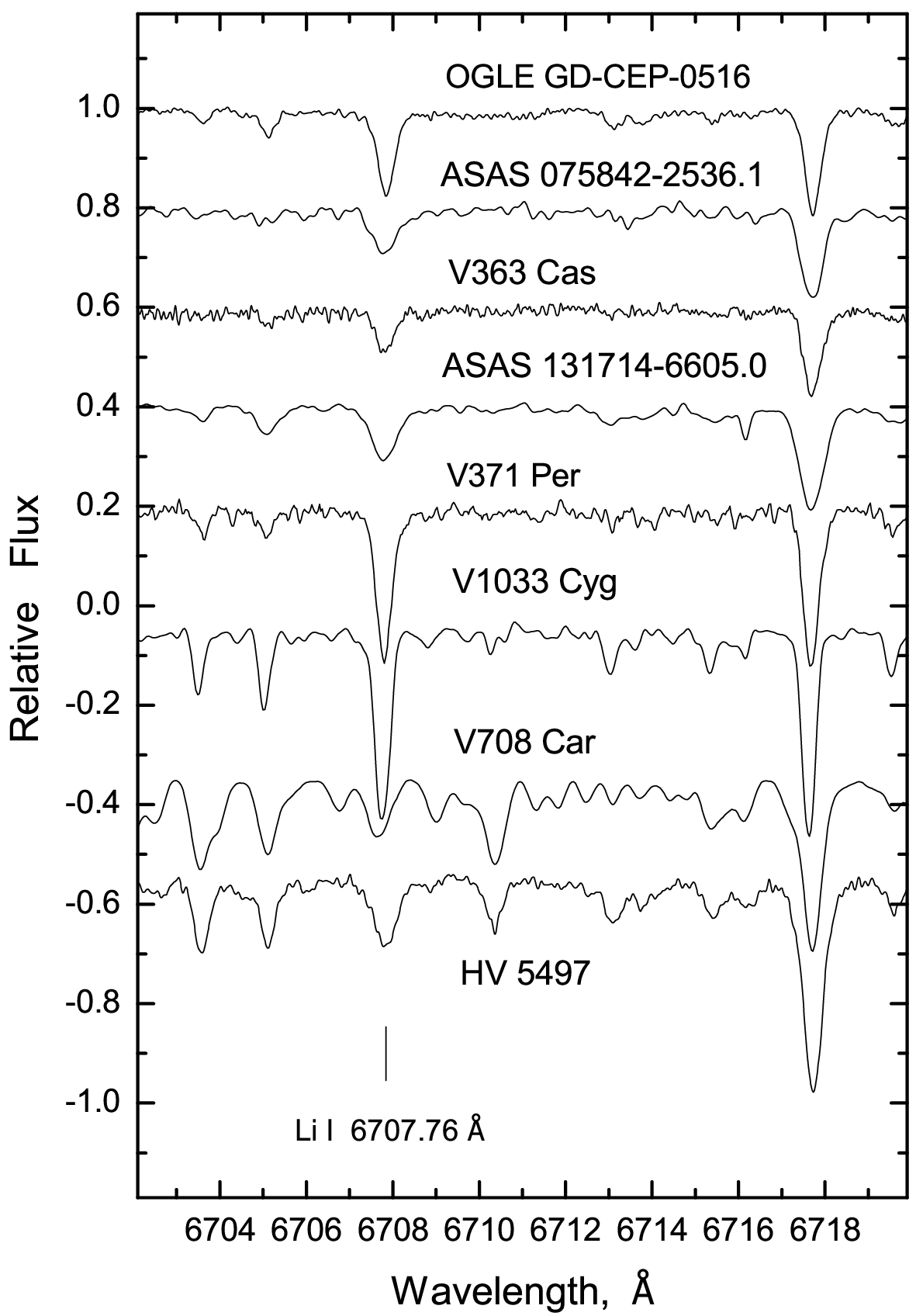}}
\caption{Lithium line in the spectrum of V708 Car and other lithium-rich Cepheids. 
}
\label{Liline}
\end{figure}

Fig. \ref{Liline} shows the Li\I\ 6707.8\,\AA\ spectral line in our program Cepheid.
High lithium abundances are very rare in supergiant stars because the lithium nuclei are destroyed by reactions with 
protons at the bottom of the convection zone. Nevertheless, some examples of lithium-rich
supergiants exist. 
For instance, the F-type supergiant HD~172365 was discovered by \cite{Luck1982} 
as a high Li abundance star. \cite{Andrievskyetal1999} found in a spectroscopic 
study that the lithium abundance in this star is $\log$~A(Li) = 2.9. These authors discussed 
the properties of this supergiant  and concluded that HD~172365 may be a post-blue straggler 
star with a mass about twice as high as the turn-off point mass for the open cluster IC~4756, 
of which HD~172365 is a member. This star could be formed by a merger process of two companions 
of a close binary system. Another supergiant, HD~174104, was also described as lithium rich
by \cite{Luck1982}.  

While non-variable supergiants are  situated near or outside the instability strip, some
pulsating yellow supergiant Cepheids also show an increased lithium abundance. The first lithium-rich 
Cepheid HV~5497 from the Large Magellanic Cloud was described by \cite{LuckLambert1992}. 
\cite{LuckLambert2011} reported increased lithium level in the Cepheid V1033~Cyg, and \cite{Kovtyukhetal2016}
indicated that the lithium abundance in the Cepheid V371~Per is high. Three Cepheids with a high lithium abundance were described 
by \cite{Kovtyukhetal2019} (namely, ASAS 075842-2536.1 and ASAS 131714-6605.0), and OGLE GD-CEP-0516 (\citealt{Kovtyukh2023}). 
All have solar metallicity and $\log$~A(Li) of about 3.0. Tables~\ref{Cepheids} and \ref{Supergiants} list some characteristics 
of the lithium-rich Cepheids and non-variable stars.

\begin{figure}
\resizebox{\hsize}{!}{\includegraphics{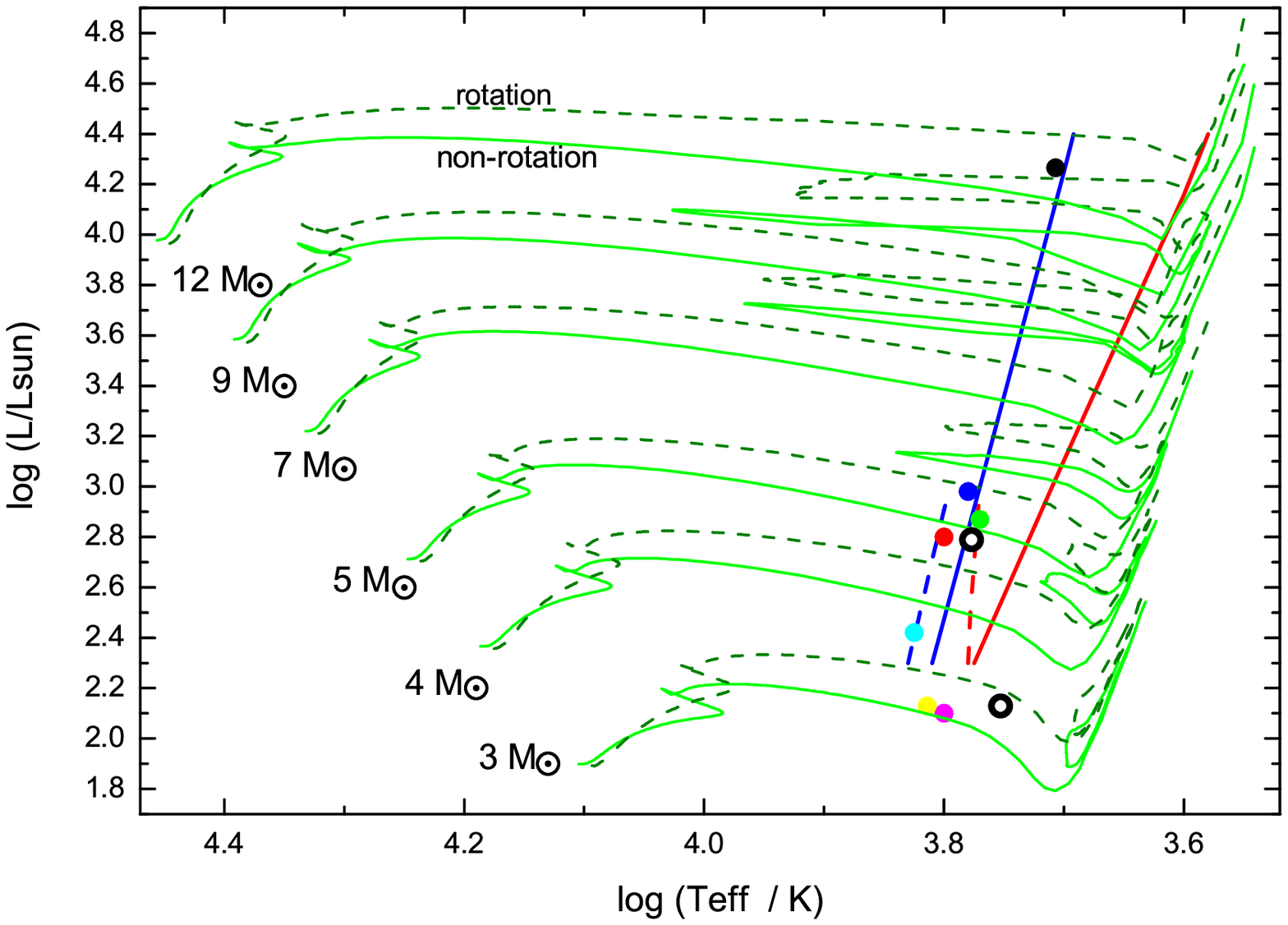}}
\caption{HR diagram. The positions of our program
Cepheid (black circle) and other lithium Cepheids described in the literature are indicated
(V371 Per, blue; V1033 Cyg, green;   ASAS 131714-6605.0, red; V363 Cas, cyan; ASAS 075842-2536.1, yellow;
 OGLE GD-CEP-0516, magenta circles; and two nonvariable supergiants, open circles). We plot the evolutionary tracks from  
\cite{Ekstrometal2012} for 3--12 solar masses with and without rotation.  The solid and dashed lines denote the blue 
and red boundaries of the instability strip for the usual classic Cepheids and so-called s-Cepheids (first-overtone 
Cepheids have sinusoidal light curves), respectively.}
\label{HR}
\end{figure}

Fig.\,\ref{HR} shows a HR diagram in which the position of our program Cepheid is
indicated, as well as the positions of other lithium-rich Cepheids and non-variable supergiants
discussed above. To determine the luminosity, the period-luminosity relation was used, and metallicity was taken into account 
(\citealt{Breuval2022}). It should be noted that all Cepheids (within the measurement error) are situated near the edges 
of the instability strip. We have to remark that \cite{Berdnikov2010} argued that V708~Car is a Cepheid that is crossing 
the instability strip for the third time. If we accept this supposition, then it is very difficult to explain the high lithium abundance 
in this star because in this case, it should already have experienced the first and second dredge-up episodes.
We leave this question open for now.

It is interesting to note that the lithium-rich Cepheids include five double-mode pulsators. Double-mode Cepheids form 
a rare group of classical Cepheids. They simultaneously pulsate in either in the second P$_{2}$ and first P$_{1}$ overtones or in the first overtone 
and fundamental mode P$_{0}$ (Table~\ref{Cepheids}).

\begin{table*}
\begin{center}
\caption{Parameters of the lithium-rich classical Cepheids.}
\label{Cepheids}
\begin{tabular}{lllrrcrrl}
\hline\hline
   Star            &  Mode & Period     &$<$V$>$&   \Rg   &    $d$  &   $l$   &    $b$   &Remarks    \\ 
                   &       & days     & mag   &  kpc    &  pc   &  deg  &  deg   &           \\ 
\hline
  Galaxy:  &&&&&&&& \\
OGLE GD-CEP-0516   & P2/P1 &0.394959 (P1)  & 12.666&  7.87   & 2720  & 285.50& $-$1.45 & \cite{Kovtyukh2023} \\ 
ASAS 075842-2536.1 & P2/P1 &0.41013 (P1)  & 12.260 &  9.03   & 2100  & 243.55&  +2.35 &\cite{Kovtyukhetal2019} \\
V363 Cas           & P2/P1 &0.546597 (P1) & 10.550 &  8.76   & 1155  & 118.46& $-$2.22 &\cite{Catanzaroetal2020}\\
ASAS 131714-6605.0 & P2/P1 &0.913165 (P1) & 11.820 &  6.85   & 2200  & 305.86& $-$3.64 &\cite{Kovtyukhetal2019} \\ 
V371 Per           & P1/P0 &1.738 (P0)    & 10.930 & 10.61   & 3200  & 146.02&$-$14.65 &\cite{Kovtyukhetal2016} \\
V1033 Cyg          &   P0  &4.9375119 & 13.027 &  7.52   & 3429  &  69.94&  +0.49 &\cite{LuckLambert2011}   \\ 
V708 Car           &   P0  &51.403084 & 12.098 &  8.24   & 4378  & 284.34& $-$2.43 & this paper              \\  
  LMC:   & & & &&&&& \\
HV 5497            &   P0  &99.156076 & 11.930 &      -- &    -- & 277.24& --36.17&\cite{LuckLambert1992}    \\
\hline
\end{tabular}
\end{center}
\begin{itemize}
\item[] {Remarks:}
\Rg\ is the galactocentric distance,  and
d is the heliocentric distance.
\end{itemize}
\end{table*}

\begin{table*}
\begin{center}
\caption{Parameters of the lithium-rich non-variable supergiants.}
\label{Supergiants}
\begin{tabular}{lllrcccclrll}
\hline\hline
HD&Sp&V&B--V&\vsini &\Mv&\Teff&\logg&\Vt  &[Fe/H]&(Li/H)& Remarks \\
 &  &mag& mag&\kms& mag& K&  & \kms& dex &  &    \\
\hline
\hline
174104 & G0Ib&8.37 & 0.690 & 50.0 & --0.69&5854 & 2.6 &5.0& 0.16 &3.46 &\cite{Luck1982} \\
172365&F8II&6.36&0.735&67.3&--2.35&5978&1.3 &7.5&--0.07&2.90& \cite{Luck1982},\\
&&&&&&&&&&&\cite{Andrievskyetal1999} \\ 
\hline
\end{tabular}
\end{center}
\end{table*}
                                                           
\section{Chemical composition peculiarities of V708 Car and their possible origins}

\begin{figure}
\resizebox{\hsize}{!}{\includegraphics{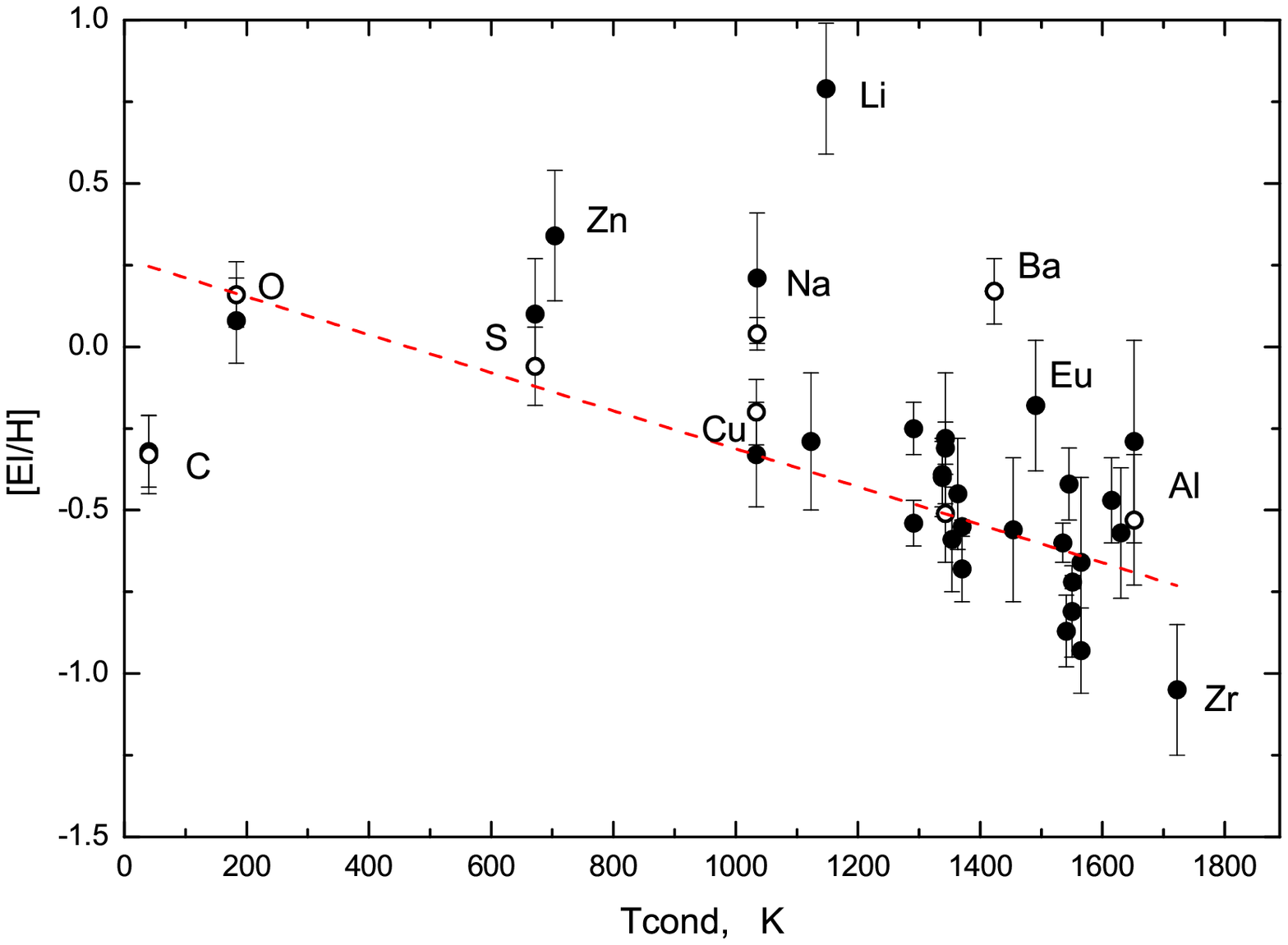}}
\caption{Element abundances relative to solar in V708~Car as a function of the condensation temperature 
(LTE is shown as filled circles,  and NLTE is shown as open circles). The correlation coefficient is 0.593 (0.669 without Li\I).}
\label{Ab_Tcond}
\end{figure}

Fig.\,\ref{Ab_Tcond} shows the results of our abundance measurements relative to solar values for 27 chemical elements 
as a function of the their condensation temperature. The existing dependence can be traced even by eye. The abundances of elements 
whose condensation temperature is below 800\,K (volatiles) are close to the solar values, and the 
abundances of elements with a condensation temperature higher than 1000\,K (refractory elements) are mainly 
underabundant. It is interesting to note that the volatile lithophile elements (Na, Zn) show higher abundances 
compared to other volatile elements (C, N, O, including volatile siderophile S). This is in accordance 
with the abundances found in chondrites (\citealt{Woodetal2019}). According to Wood et al. (2019), lithophile refractory elements (e.g. Mg, Al, Si, Ca, Sc, Ti,  Zr, rare-earth elements, and Eu) 
in chondrites are much more abundant than volatile elements, while refractory siderophile elements (e.g. V, Cr, Mn, Fe, Ni, Co, and Cu) 
have similar abundances. In the case of our program star, the former and latter groups of elements show much lower abundances than 
the volatiles. This testifies that some processes occurred in their environment that caused the apparent deficiency of these elements.    

According to the definition of \cite{Woodetal2019}, volatility is conventionally defined geochemically in a very broad sense 
as being  related to the temperature at which a specific element would condense from a gas of solar composition. Thus, volatile elements 
are defined as those that condense at relatively low temperature (e.g., less than 1100\, K) from the putative solar gas while refractory, involatile 
elements condense at higher temperatures.

One of the reasons for the dependence of the abundances of elements on their condensation 
temperature can be understood considering the chemical anomalies of the $\lambda$~Boo-type stars,
W~Vir stars, and post-asymptotic giant branch (post-AGB) stars. For all these objects, the elemental abundance 
distribution versus condensation temperature is qualitatively similar and follows the trend described above. The mechanism 
that explains the chemical peculiarities of these stars has been discussed in several papers.
It is based on the assumption that the stellar atmospheres were contaminated by the material of a circumstellar
shell. To reach this specific chemical composition, we need to assume that the gas of a circumstellar
envelope experienced a separation of dust and gas. Refractory elements can form dust particles at a
quite high temperature, and then they can be swept out of the envelope by radiative pressure from the star. 
The volatile elements stay in the gaseous fraction of the envelope, and the subsequent selective accretion 
of these elements by the star leads to the specific chemical property of the envelope. The atmosphere of the 
star becomes deficient in refractory elements and has a normal abundance of volatile elements. In the case of 
Cepheids, this is true at least for the abundances of elements such as oxygen, sulfur, and zinc because the abundances 
of carbon and nitrogen can be altered during the first dredge-up episode (carbon is deficient, while
nitrogen is overabundant).

For example, the mechanism of the dust--gas separation was applied to explain the chemical peculiarities seen
in $\lambda$~Boo-type stars (see, e.g. \citealt{VennLambert1990}, \citealt{Charbonneau1991},
\citealt{Charbonneau1993}, \citealt{Stuerenburg1993}, \citealt{Andrievsky1997}, \citealt{AndrievskyPaunzen2000}, 
\citealt{Andrievsky2006}, \citealt{VennLambert2008}). Sodium has a rather high condensation temperature, but is 
often overabundant (like in our program star, as can be seen from Fig.\,\ref{Ab_Tcond}). \cite{Andrievsky2006} explained 
this observed fact in $\lambda$~Boo-type stars by the specific ionization regime of the sodium atoms that is established in 
the circumstellar envelope. The same mechanism might be valid for barium, which has a fairly high
condensation temperature, but is apparently overabundant. Sodium and barium both have very similar first ionisation
potentials of about 5~eV. However, it is unclear why many other chemical elements with a first ionisation potential slightly 
higher than that of sodium and barium are deficient.

$\lambda$~Boo-type stars are A--F stars. However, the process of dust--gas 
separation can be also efficient for G stars. For example, \cite{Maasetal2007}, \cite{Kovtyukhetal2018} found 
similar trends for type II Cepheids. A similar depletion pattern was also described by \cite{Oomenetal2019} for 
several post-AGB stars. These authors modelled the re-accretion of gas from a dusty circumstellar disc.

As a consequence of the dust--gas separation (\citealt{Yushchenkoetal2022}), the dust particles that are swept out of the envelope can form coagulated planetesimals. For instance, 
\cite{Melendezetal2009} and \cite{Ramirezetal2009} showed that our Sun has a small deficiency of refractory 
elements compared to solar twins, and this may indicate that some fraction of the material containing refractory 
elements was bound in the planets. 

An indirect indication of the presence of a circumstellar envelope (or extended atmosphere) for classical Cepheids exists 
in the literature. However, the origin of these structures as envelopes or extended atmospheres in these stars is unknown.
One solution of the problem could be permanent mass loss either at the stage of a red supergiant or at the later stage 
of a Cepheid. A similar mechanism was used by \cite{vanLoonetal2005}, who showed that the mass-loss rate increases with 
increasing luminosity of the star (we recall that our program star is very luminous; see the Introduction).
It is interesting to note that according to \cite{Kervellaetal2019}, the long-period Cepheid RS~Pup 
is embedded in the dusty circumstellar nebula, but its origin is unclear.

\begin{figure}
\resizebox{\hsize}{!}{\includegraphics{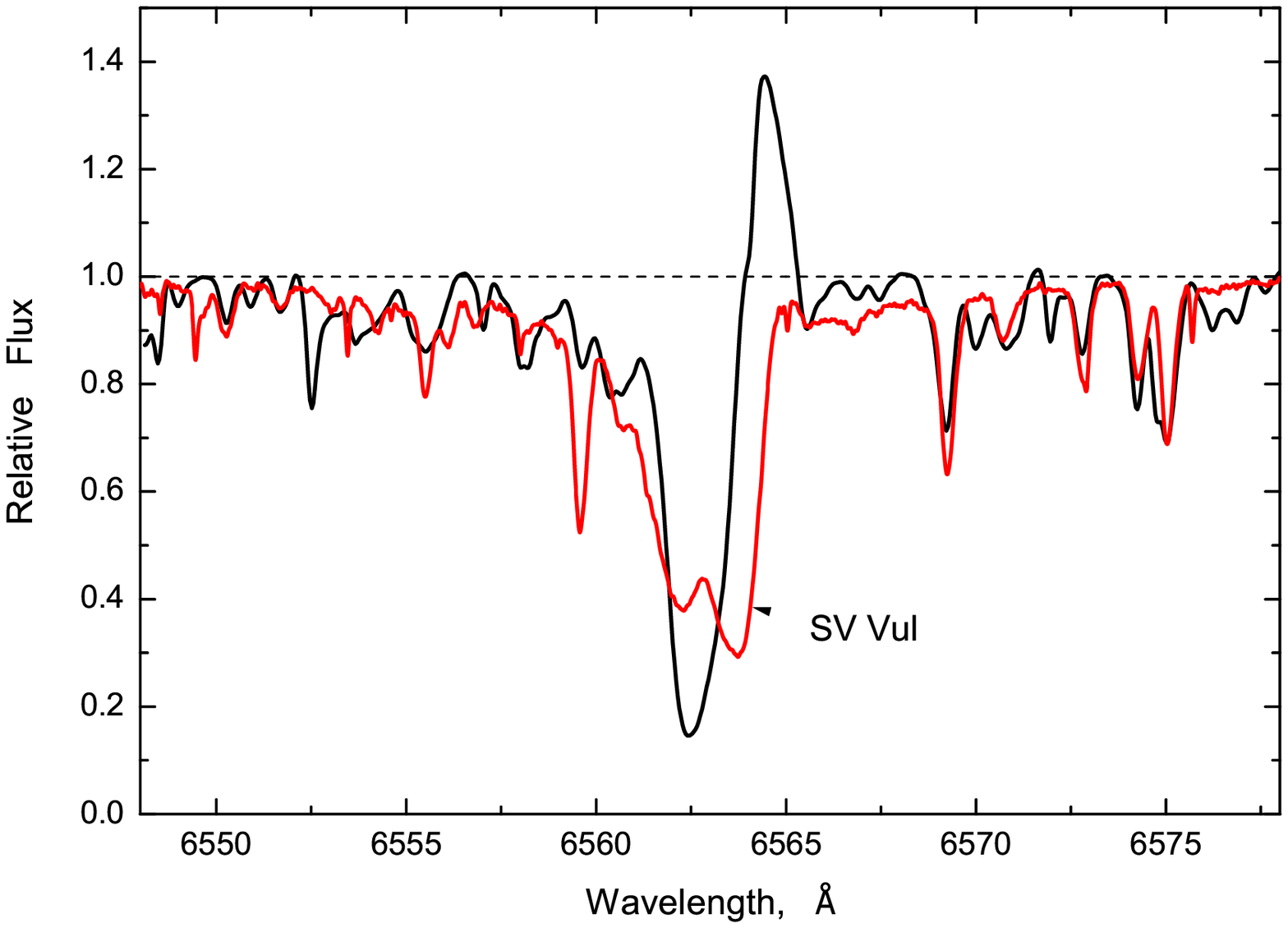}}
\caption{H$\alpha$ profile in the spectrum of V708 Car (solid black line). For comparison, the spectrum of the 
Cepheid SV~Vul ($P=44.89$~d, solid red line) is shown, whose parameters are similar to those of V708~Car, 
and which was observed in the same pulsation phase 0.22.}
\label{Halpha}
\end{figure}

Our program star has strong H$\alpha$ emission (see Fig.\,\ref{Halpha}). This emission may serve 
as indirect evidence of the gaseous envelope surrounding V708~Car. The 2MASS all-sky catalogue (\citealt{Cutrietal2003}) 
lists the magnitudes of V708~Car in different bands. The star appears very bright in the H and K bands (about 6~mag), but is quite faint 
in the V band (about 12~mag).  

Since V708~Car is situated at the blue edge of the Cepheid instability strip (Fig.\,\ref{HR}), convection probably
had no time to erase atmospheric chemical peculiarities. Its pulsation amplitude is small (0.55~mag in V band; \citealt{Berdnikov2010}), 
and this could be additional evidence that this star is just at the border of the instability strip.
However, we stress again that \cite{Berdnikov2010} considered this star to be at its third crossing 
of the instability strip.

We lack information about the properties of the present-day envelope of V708 Car, 
including temperature  or the density-pressure distributions. This is probably not very important, however, because the mechanism of the gas--dust separation described above could have 
occurred earlier, at the MS stage, or immediately after it.  The position of V708 Car in 
the HR diagram (Fig.\,\ref{HR}) indicates that its mass is rather high, so that it could be formed as a result of the 
evolution of two stars. For example, if two MS intermediate-mass B stars (with masses of about 7 and 5 solar masses, e.g.) 
were to undergo a common evolution in a binary system with mass  transfer,  the result  of this process would be a
B (super)giant star with a mass of about 12 solar masses (and with a merged or unmerged white dwarf that previously was the primary component). 
It is possible that during this evolution, a common envelope forms. Since B stars do not have convective 
zones in their outer atmosphere, their atmospheres are affected by the gas--dust  separation in the envelope and 
will retain a chemical anomaly for a long time. This situation will not change after  the further excursion of the star that
formed in this way to the instability strip. We probably observe V708 Car in this stage. 

We cannot reproduce the physical conditions in the relic envelope of V708 Car. We therefore applied the widely 
used condensation temperatures provided by \cite{Lodders2003}. They are given for a pressure of 10$^{-4}$ bar. 
According to the Gay-Lussac law, pressure is linearly proportional to temperature. For any local value 
of pressure in the envelope, the overall picture of the condensation temperatures for different species therefore 
scales proportionally.

\section{Conclusion}

We discovered and described the lithium-rich Cepheid V708~Car ($\log$~A(Li) = 1.95), which together with 
 six already known lithium Cepheids and two lithium-rich non-variable supergiants forms 
a small group of lithium-rich yellow supergiants. Our program star also exhibits an anomalous chemical
composition. The measured abundances of the chemical elements show a clear correlation with their condensation
temperature. We explained this phenomenon as the result of dust--gas separation in the 
envelope (or extended atmosphere) of V708~Car. Through this mechanism, refractory elements form 
dust grains in the envelope that are swept out of the envelope by the radiative pressure of 
the star. The gaseous fraction, enriched in volatile elements, is accreted by the star, and eventually,
its atmosphere acquires the observed peculiarity.  

We probably caught the star at a very short-term evolutionary stage of interaction with the shell.


\begin{acknowledgements}
SMA would like to acknowledge the support from the ESO Scientific Visitor Programme. The authors thank the anonymous referee for their careful reading of the paper and for their comments, which improved the clarity of our manuscript.
All observations reported in this paper were obtained within observational program 2016-1-MLT-002
at the Southern African Large Telescope (SALT).
AK acknowledges the National Research Foundation (NRF) of South Africa.
This research has made use of the SIMBAD database, operated at CDS, Strasbourg, France. 
\end{acknowledgements}
\bibliographystyle{aa}
\bibliography{aa}
\end{document}